\documentclass[journal,12pt,onecolumn,draftclsnofoot,]{IEEEtran}
\usepackage{graphicx}
\usepackage{subcaption}
\usepackage{amsmath}
\usepackage{amssymb}
\usepackage{mathrsfs}
\usepackage{stfloats}
\usepackage{dsfont}
\usepackage{setspace}
\usepackage{epstopdf}
\usepackage{epsfig}
\usepackage{color}
\usepackage{epsfig}
\usepackage{tabularx,ragged2e,booktabs,caption}
\usepackage{balance}
\usepackage[printwatermark]{xwatermark}
\usepackage{xcolor}
\usepackage{graphicx}
\usepackage{lipsum}

\ifCLASSINFOpdf
\else
\fi
\begin{document}
%
\title{Evaluating the Performance of eMTC and NB-IoT for Smart City Applications}

\author{Mohieddine~El Soussi, Pouria~Zand, Frank Pasveer and~Guido~Dolmans\\
\IEEEauthorblockA{Holst Centre/imec, Eindhoven, The Netherlands\\
e-mail:\{mohieddine.elsoussi, pouria.zand, frank.pasveer and guido.dolmans\}@imec-nl.nl}}


\maketitle

\begin{abstract}
Low power wide area network (LPWAN) is a wireless telecommunication network that is designed for interconnecting devices with low bitrate focusing on long range and power efficiency. In this paper, we study two recent technologies built from existing Long-Term Evolution (LTE) functionalities: Enhanced machine type communications (eMTC) and Narrow band internet of things (NB-IoT). These technologies are designed to coexist with existing LTE infrastructure, spectrum, and devices. We first briefly introduce both systems and then compare their performance in terms of energy consumption, latency and scalability. We introduce a model for calculating the energy consumption and study the effect of clock drift and propose a method to overcome it. We also propose a model for analytically evaluating the latency and the maximum number of devices in a network. Furthermore, we implement the main functionality of both technologies and simulate the end-to-end latency and maximum number of devices in a discrete-event network simulator NS-3. Numerical results show that 8 years battery life time can be achieved by both technologies in a poor coverage scenario and that depending on the coverage conditions and data length, one technology consumes less energy than the other. The results also show that eMTC can serve more devices in a network than NB-IoT, while providing a latency that is 10 times lower.

\end{abstract}

%

%
\IEEEpeerreviewmaketitle

\section{Introduction}
The internet of things (IoT) interconnects a massive number of devices, e.g., machines, vehicles and sensors, to the internet and exploits the data that is generated.  The IoT world is growing fast, from 2 billion objects in 2006 to a projected 200 billion by 2020 \cite{Intel17}. IoT is expected to create steady growth in the economy since these smart objects give vital data to track inventory, manage machines, increase efficiency, save costs, and even save lives. 

While many IoT devices will be served by short-range radio technologies that operate on an unlicensed spectrum such as WiFi, ZigBee and Bluetooth, a significant proportion will be enabled by wide area networks (WANs) \cite{LEIOT17}. Low power wide area network (LPWAN) is a type of wireless telecommunication intended for wireless battery operated things and designed to allow long range communications at a low bit rate. Currently, there are two connectivity tracks for LPWAN, one operating on an unlicensed spectrum such as SigFox and LoRa \cite{RKS17} and one operating on a licensed spectrum such as Cellular IoT.

The third generation partnership project (3GPP) has introduced two Cellular IoT technologies based on Long Term Evolution (LTE), namely: eMTC (enhanced Machine Type Communications) or CAT-M1 \cite{TS36300}, and Narrowband Internet of Things (NB-IoT) or CAT-NB1 \cite{TR45820}. These systems are designed to coexist with existing LTE infrastructure, spectrum, and devices. eMTC targets applications such as VoLTE (Voice over Long-Term Evolution), tracking devices and objects that require mobility, high data rate and low power consumption with wide area coverage. NB-IoT targets applications such as control equipment, sensors, and meters that require low complexity and low power consumption with wide area coverage.

In smart cities, a massive number of IoT devices will be deployed for several use cases. Key challenges to enabling a large-scale uptake of massive IoT include: device costs, battery life, scalability, latency and coverage. In this paper, we consider some of these challenges while studying the energy consumption, the latency and scalability for both technologies. 

In order to ensure long battery lifetime and to reduce energy consumption, these technologies are enabled with two power saving features: extended discontinuous reception (eDRX) and power saving mode (PSM) \cite{TS23682}. eDRX is a mechanism that enables the device to switch off part of its circuitry to save power. An eDRX cycle consists of an ``On Duration" during which the device checks for paging and an ``eDRX period" during which the device is in sleep mode. This feature is useful for device-terminated applications, e.g., smart grid. PSM is a low-power mode that allows the device to skip the periodic page monitoring cycles, allowing the device to sleep for longer. However, as a result the device becomes unreachable. It is therefore best utilized by device-originated or scheduled applications, e.g., smart metering.   

The energy consumption of both eMTC and NB-IoT devices has been studied in, for example, \cite{LKMSH16, RVMG16}, where a highly accurate expensive clock has been considered for deep sleep periods. However, since another key requirement of these technologies is low cost, it is beneficial to consider a low cost, low power clock for deep sleep periods at the expense of having a clock drift. Clock drift causes the device to loose its time and frequency synchronization. Furthermore, depending on the length of the sleep period and the amount of time and frequency drift, more time is needed to synchronize. Hence, the device must wake up earlier to synchronize and to not miss its scheduled transmission or reception. This increases the energy consumption of the device since the accurate clock is turned on for additional duration. In this paper, we study this problem and reduce the energy consumption by allowing the device to wake up several times during the sleep period to synchronize. We formulate an optimization problem that aims to evaluate the sleep duration and the number of waking up times in order to minimize the total consumed energy during the sleep period.    

Other key requirements for these technologies are scalability and latency. In general, IoT devices are deployed on a large scale and in order to a priori study the network performance, expensive test-beds are required. Network simulators such as NS-3 offer an alternative solution to those expensive test-beds to test and evaluate the performance of large-scale networks. Based on the existing LTE module in NS-3, we implement eMTC and NB-IoT modules in NS-3, where we adapt both the physical (PHY) layer and the media access control (MAC) layer according to eMTC and NB-IoT standards. In the PHY layer, we modify the PHY error model by introducing the link-level results of the convolution code for different modulation and coding schemes (MCSs). In addition, we limit the number of resource blocks (RBs) to 6 RBs and 1 RB for eMTC and NB-IoT, respectively. In the MAC layer, we include switching and cross subframe delays for the control and data channels according to 3GPP specifications. The latency based on system level has been studied and evaluated in \cite{RVMG16} and \cite{R1156006}. In this paper, in addition to system level evaluation, the latency and scalability are also evaluated at the network level using NS-3. Thus, we estimate the end-to-end latency from sensors/actuators to the cloud as well as the maximum number of devices that can be served by a cell. In addition to the simulation results, we formulate the latency and the maximum number of sensors based on the channel quality, the maximum coupling loss (MCL) and the transmission period.

Through simulation results, we show that at least 8 years battery life time can be achieved by both technologies in a poor coverage scenario taking into account the clock drift. We also show that depending on the coverage conditions and data length, one technology consumes less energy than the other and that eMTC can serve more devices in a network than NB-IoT, while providing a latency that is 10 times lower.
We also formulate the data transmission delay and the maximum number of UEs per cell or network and demonstrate that the results derived from analytical formulations correspond to the results obtained from the network simulator.

\section{System Design Overview}
In this section, we briefly provide an overview of eMTC and NB-IoT with a focus on the key aspects where they deviate from LTE.

eMTC is an evolution of LTE optimized for IoT. It has introduced a set of physical layer features that aim to reduce cost and power consumption while extending coverage \cite{RVXWBBTY16}. eMTC devices operate with a bandwidth of 1.08MHz (6 LTE Physical Resource Blocks (PRBs)) for the transmission and reception of physical channels and signals. The downlink of eMTC is based on orthogonal frequency division multiplexing (OFDM) scheme with 15kHz subcarrier spacing as in LTE. In the uplink, eMTC also uses the same numerology as LTE. The transmission is based on single carrier frequency division multiple access (SC-FDMA) with 15kHz subcarrier spacing. The reader can refer to \cite{RVXWBBTY16} for more details.

NB-IoT is a new 3GPP radio-access technology built from LTE functionalities. It is not backward compatible with existing 3GPP devices. It is, nevertheless, designed to coexist with LTE. NB-IoT occupies a frequency band of 180kHz, which is further decreased compared to eMTC, and corresponds to one PRB in LTE transmission. This narrowband bandwidth allows the device complexity to be further reduced at the expense of lower data rate. The downlink of NB-IoT is based on OFDM with 15kHz subcarrier spacing as in LTE (the PRB contains 12 subcarriers). In the uplink, NB-IoT also uses the same numerology as LTE, however NB-IoT supports not only multi-tone but also single-tone transmissions. A single-tone transmission supports 3.75kHz and 15kHz subcarrier spacing. The 3.75kHz numerology uses 2ms slot duration instead of 0.5ms, and the PRB contains 48 subcarriers instead of 12 subcarriers to remain compatible with the LTE numerology. The 15kHz numerology is identical to LTE. In the multi-tone transmissions, 3, 6, or 12 tones with 15kHz subcarrier spacing can be used and the transmission is based on SC-FDMA scheme \cite{RMZRK16}. The reader can refer to \cite{WLAGSBBR17} for more details.
\section{Power Consumption Analysis}

\begin{figure*}[ht!]
  \begin{center}
  \includegraphics[width=.9\linewidth]{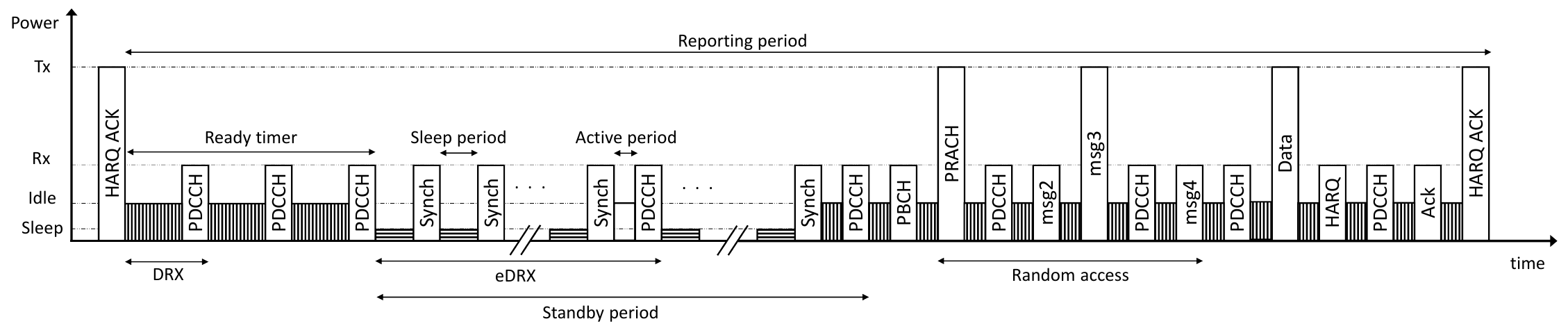}
  \end{center}
  \caption{Different states of the power model and the resulting power levels.}
	\label{Power_model}
\vspace{-.5cm}
\end{figure*}

Low power consumption is a key requirement for these technologies to ensure years of battery lifetime. In Figure \ref{Power_model}, we present a model for calculating the energy consumption of an LTE device as described in \cite{TR45820}. The model shows the different events performed by the device to transmit a packet and the corresponding power consumption of each event. 

The model considers four different operating states: transmission state, receiving state, idle state and sleeping state. In this model, we assume that the device transmits with a reporting interval equal to $t_{tot}$. To transmit a report, the device needs to obtain downlink synchronization (Synch period), acquire the system information (PBCH period) and achieve the uplink synchronization (random access period). Once the random access procedure is completed, the device waits for an uplink grant on the physical downlink control channel (PDCCH), transmits the report and then waits for the network and application acknowledgement. The device monitors the control channel according to its DRX cycle until the ready timer expires, and then it either continues to monitor the control channel according to its idle eDRX cycle or sleeps for a longer period according to its PSM (Standby period). 

During the eDRX cycle, the receiver is switched on for a duration of $t_{\text{PDCCH}}$ to check for paging and then goes back to sleep. During the sleep period, most of the device circuitries are switched off, including the high accurate crystal clock and only the very low power clock is enabled. Unfortunately, this clock is inaccurate and causes the device to loose its time and frequency synchronization. As a result, the device might wake up late and miss the scheduled paging, transmission or reception. Hereafter, we assume that the error due to clock drift only depends on the length of the sleep period. Hence, the device's circuitry must be switched on $m*t_{\text{sleep}}$ earlier to synchronize with the network where $m$ is the fractional error. This increases the power consumption and thus decreases the lifetime of the battery. We should note that the accurate clock is switched on during the synchronization period and it is switched off only during the sleep period. In order to reduce the power consumption, the receiver, during the off period of the eDRX cycle, wakes up and then goes back to sleep several times to obtain the time and frequency synchronization. Following this, the high accurate clock stays switched on for a duration of $t_{\text{act}}=m*t_{\text{sleep}}$ after the last synchronization and before receiving the control channel (PDCCH) as shown in Figure \ref{Power_model}. 
In order to minimize the total energy consumed during an eDRX period, the sleeping duration should be large, the number of waking up should be small and the active time should be also small. However, this is not achievable since the active time is proportional to the sleeping time. To overcome this problem, the device will have different sleeping duration values. The first sleeping duration is the largest to reduce the number of waking up for synchronizing and the following ones are smaller so that $t_{\text{act}}$ is minimized. Hence, the optimization problem to reduce the energy consumption during an eDRX period can be stated as:
\begin{subequations}\label{OP}
\begin{align}
    \min_{K,\ t_{\text{act}}^K,\ t_{\text{sleep}}^k} \: \: & KP_{\text{Rx}}t_{\text{synch}} + \sum_{k=1}^KP_{\text{sleep}}t_\text{sleep}^k + P_{\text{Idle}}t_{\text{act}}^K \label{eq_opt_1a}\\
    \text{s.t.} \quad & t_{\text{eDRX}} - t_{\text{PDCCH}} - Kt_{\text{synch}} - \sum_{k=1}^Kt_{\text{sleep}}^k = t_{\text{act}}^K \label{eq_opt_1b}\\
    & t_{\text{act}}^K = mt_{\text{sleep}}^K  \label{eq_opt_1c}\\
    & t_{\text{sleep}}^k > 0, \quad \forall k,
\end{align}
\end{subequations}
where $t_{\text{synch}}$ is the synchronization time, $t_{\text{PDCCH}}$ is the time to check for paging, $K$ is the number of cycles or iterations, $t_{\text{sleep}}^k$ is the sleeping time at cycle $k$, $t_{\text{act}}^K$ is the active time during the last cycle and $P_{\text{Rx}}$, $P_{\text{sleep}}$ and $P_{\text{Idle}}$ are the power value for the receiving, sleeping and idle states, respectively.

The optimization problem is a mixed integer optimization problem and it is hard to find $K$, $\{t_{\text{sleep}}^k\}_{k=1}^K$ and $t_{\text{act}}^K$ jointly in a reasonable time. Therefore, we propose an iterative optimization, where we find appropriate $\{t_{\text{sleep}}^k\}_{k=1}^K$ and $t_{\text{act}}^K$, and $K$, alternately. More specifically, for a given value of $K$, we can compute $\{t_{\text{sleep}}^k\}_{k=1}^K$ and $t_{\text{act}}^K$ using \eqref{OP}. It can be easily seen from \eqref{eq_opt_1a} that the energy consumption is minimized for small $K$ and $t_{\text{act}}^K$. Thus, the optimization starts by setting $K=1$ and optimizing over $\{t_{\text{sleep}}^k\}_{k=1}^K$ and $t_{\text{act}}^K$. Next, we increment $K$ by one and optimize again over $t_{\text{sleep}}^K$ and $t_{\text{act}}^K$ given the values of $\{t_{\text{sleep}}^k\}_{k=1}^{K-1}$ from the previous iterations. The iterative process terminates if either of these conditions hold: $t_{\text{act}}^K<t_{\text{synch}}$ or $P_{\text{Rx}}t_{\text{synch}} + P_{\text{sleep}}t_\text{sleep}^K + P_{\text{Idle}}t_{\text{act}}^{K}> P_{\text{Idle}}t_{\text{act}}^{K-1}$. For the first condition, the minimum energy consumption is attained using $\{t_{\text{sleep}}^k\}_{k=1}^K$ and $t_{\text{act}}^K$ and for the latter case the minimum energy consumption is attained using $\{t_{\text{sleep}}^k\}_{k=1}^{K-1}$ and $t_{\text{act}}^{K-1}$.

The above optimization can also be used for PSM by simply replacing $t_{\text{eDRX}}$ with $t_{\text{PSM}}$ and equating $t_{\text{PDCCH}}$ to 0. The main difference is that during the standby period there exist multiple eDRX and only one PSM.

\vspace{-.2cm}
\section{Latency and Scalability Model}
\label{Sec_Data_communication_delay}

In this section, we introduce a simple model to formulate the data transmission delay and the maximum number of users per cell.   

\begin{figure}[ht!]
  \begin{center}
  \includegraphics[width=.9\linewidth]{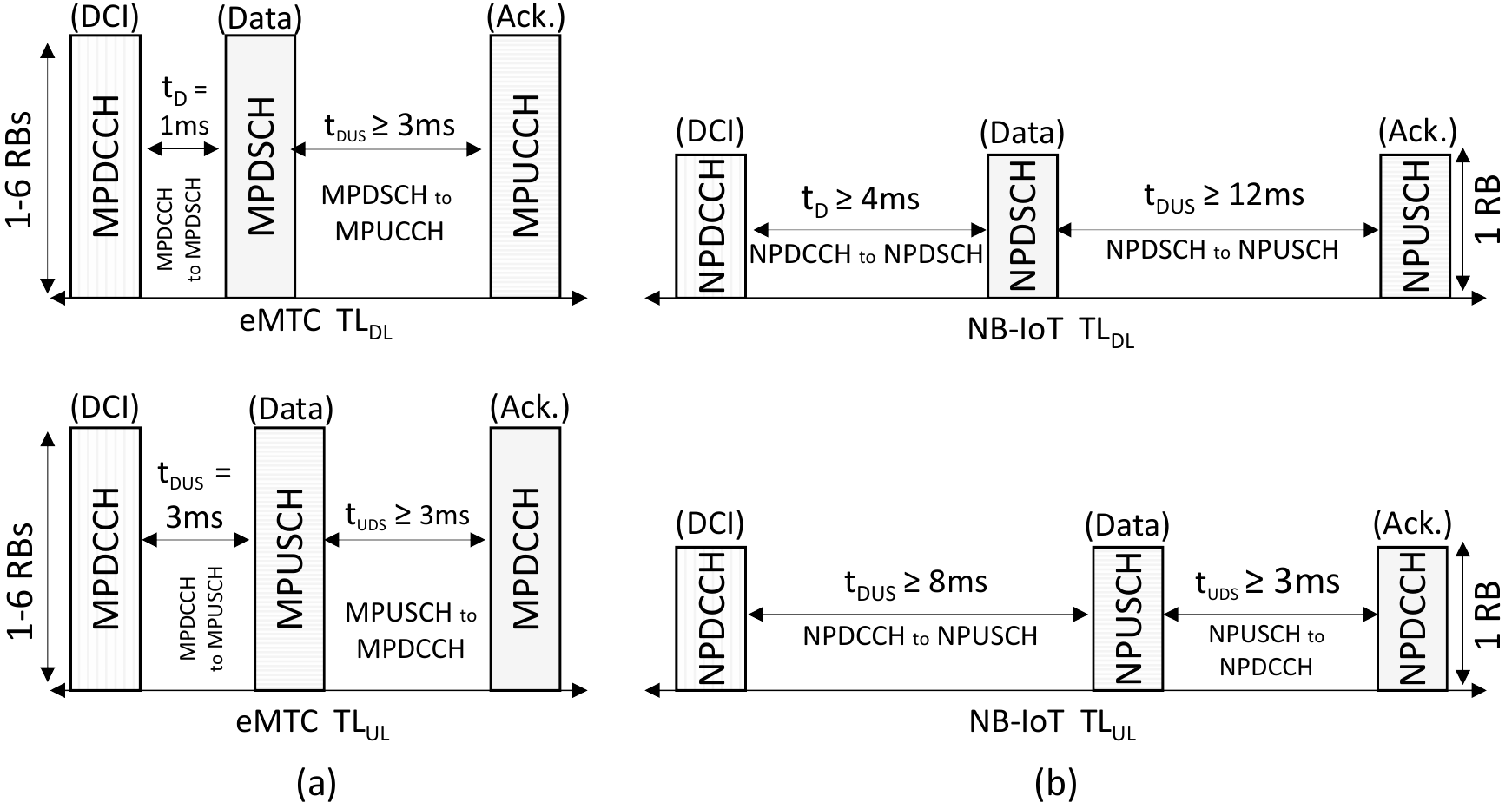}
  \end{center}
  \caption{Transmission latency for (a) eMTC and (b) NB-IoT on the downlink and uplink.}
	\label{NB-IoT-eMTC-Ul-DL}
\end{figure}

Latency consists of synchronization delay, random access channel (RACH) procedure delay and data transmission delay. Herein, we only focus on formulating the data transmission delay. The synchronization and RACH procedure delays for different MCLs are provided in \cite{R1156006}. The data transmission delay per user equipment (UE) consists of the reception of downlink control information (DCI), transmission of data and transmission or reception of the acknowledgment. The data transmission delay per UE for the downlink (DL) and the uplink (UL) transmissions are given as follows
\begin{align}
\label{eqn:delayTotal}
    \text{Delay}_{\text{UE}}^i = \text{TL}_i* \left\lceil\frac{\text{DataLen}}{\text{TBS}(\text{MCS}, \text{RBU})} \right\rceil,\hspace{1mm} i \in \{\text{DL}, \text{UL}\}.
\end{align}
It can be easily seen that the transmission delay per user depends on the total number of transport blocks needed to transmit the user data, i.e., $\left\lceil\frac{\text{DataLen}}{\text{TBS}(\text{MCS}, \text{RBU})} \right\rceil$ and the transmission latency per transport block, $\text{TL}_i$, where $\text{DataLen}$ is the data size per user, TBS is the transport block size that depends on MCS and the allocated RB per user (RBU), and $\text{TL}_i$ on the UL and DL are given by
\begin{align}
\label{eqn:delayDL}
    \text{TL}_{\text{DL}} & = \text{RLDC} * t_{\text{PDCCH}} + t_{\text{D}} + \text{RLDS}*t_{\text{PDSCH}} \nonumber\\
    & + t_{\text{DUS}} + \text{RLUC}*t_{\text{ULACK}} 
\end{align}    
\begin{align}    
\label{eqn:delayUL}
    \text{TL}_{\text{UL}} & = \text{RLDC}*t_{\text{PDCCH}} + t_{\text{DUS}} + \text{RLUS}*t_{\text{PUSCH}} \nonumber\\
    & + t_{\text{UDS}} + \text{RLDC}*t_{\text{DLACK}},
\end{align}
where $\text{RLDC}$ is the number of repetitions of the downlink control channel (PDCCH). $\text{RLDS}$ and $\text{RLUS}$ are the number of repetitions of the data on the physical downlink shared channel (PDSCH) and the physical uplink shared channel (PUSCH), respectively. $t_{\text{PDCCH}}$ is the transmission time needed to transmit the control information on the PDCCH, $t_{\text{PDSCH}}$ and $t_{\text{PUSCH}}$ are the transmission times needed to transmit one transport block on PDSCH and PUSCH, respectively. $t_{\text{D}}$ is the cross subframe delay, $t_{\text{DUS}}$ and $t_{\text{UDS}}$ are the radio frequency (RF) tuning delay for switching from DL to UL and UL to DL channels, respectively. We should note that the values of RLDC, RLDS and RLUS depend on MCL. The data transmission latency for uplink and downlink are illustrated in Figures (\ref{NB-IoT-eMTC-Ul-DL}a) and (\ref{NB-IoT-eMTC-Ul-DL}b) for eMTC and NB-IoT, respectively. In these figures, the vertical axis represents the spectrum while the horizontal one represents the time.

The total data transmission delay in a cell can be formulated as 
\begin{align}\label{eqn:Totaldelay}
 \text{Delay}_{\text{T}}^i = \text{Delay}_{\text{UE}}^i*\left\lceil \frac{\text{N}_{\text{UE}}}{\left\lfloor \frac{\text{N}_{\text{RB}}}{\text{RBU}} \right\rfloor}\right\rceil,  
\end{align}
where $\text{N}_{\text{UE}}$ is the total number of users, $\text{N}_{\text{RB}}$ is the total number of RBs and $\text{RBU}$ is the number of allocated RB per user. 

We should note that $\left\lfloor \frac{\text{N}_{\text{RB}}}{\text{RBU}} \right\rfloor$ represents the number of users in a group that can concurrently transmit and $\left\lceil \frac{\text{N}_{\text{UE}}}{\left\lfloor \frac{\text{N}_{\text{RB}}}{\text{RBU}} \right\rfloor}\right\rceil$ represents the total number of groups. 

In order to evaluate the scalability of the network, we calculate the maximum number of users per cell, which can be formulated as follows
\begin{align}
\label{eqn:maxNumberUEs}
    \max \text{N}_{\text{UE}} = \left\lfloor \frac{\text{Reporting Period}}{\text{Delay}_{\text{UE}}^i} \right\rfloor * \left\lfloor \frac{\text{N}_{\text{RB}}}{\text{RBU}} \right\rfloor,
\end{align}
where $\left\lfloor \frac{\text{Reporting Period}}{\text{Delay}_{\text{UE}}^i} \right\rfloor$ represents the total number of groups of transmission that can occur during a reporting period. In \eqref{eqn:maxNumberUEs}, we assume that the users have the same reporting period.

We should note that the scheduler is able to adapt the number of repetitions and the MCS, based on the channel quality of the link or MCL, in such a way that the block error rate is minimized at the expense of increasing the transmission air-time of the packets. We should also note that the RF tuning delay between control and data channels is needed to reduce the complexity and cost of the radio module and to support longer decoding time at the modules.

Hence, the above equations can be used to roughly estimate the data transmission delays and the maximum number of users that can be served by a cell with a very low complexity.

\section{eMTC and NB-IoT implementations in NS-3}\label{Sec_Simulation}

In this section, we provide some insights into the implementation of eMTC and NB-IoT in NS-3.

NS-3 has a complete LTE module that includes EPC control and data plane, an end-to-end data-control plane protocol stack, a physical-layer data error model, radio propagation models and different MAC schedulers. Hence, based on the LTE module in NS-3, we implement eMTC and NB-IoT modules where we adapt the LTE module according to eMTC and NB-IoT standards. We modify the PHY layer, MAC layer, and the scheduler according to 3GPP specifications.

\begin{figure}[t!]
  \begin{center}
  \includegraphics[width=.9\linewidth]{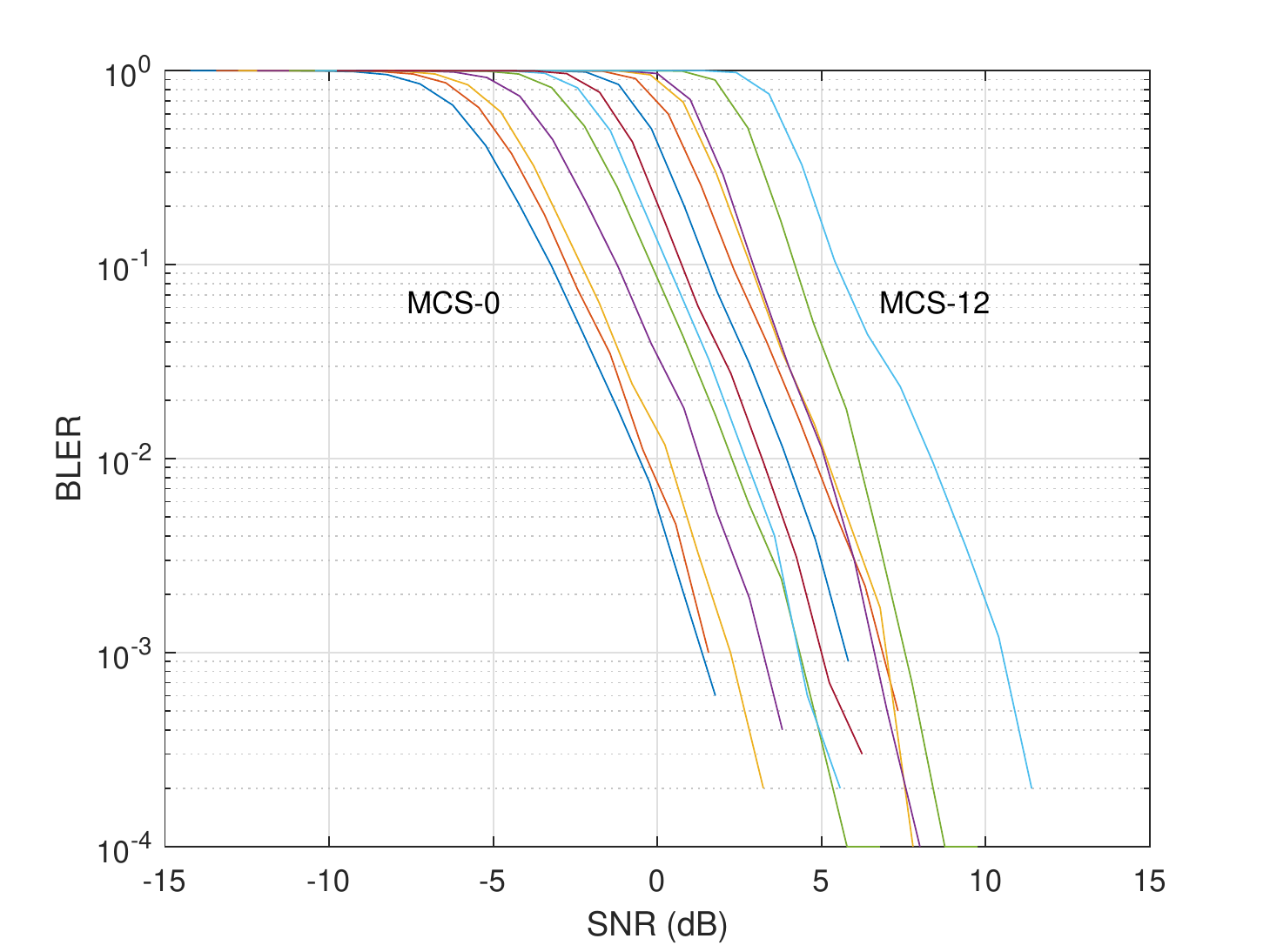}
  \end{center}
  \caption{NB-IoT BLER curves used in NS-3}
	\label{BLER-NBiot}
\vspace{-.5cm}
\end{figure}

In NS-3, a link abstraction model is used to provide an accurate link performance metric at a low computational cost \cite{MMRBZ12}. This model is based on link-level results obtained under the configuration of the PHY-layer turbo encoder in terms of code block length and MCS. In order to evaluate the PHY layer performance of downlink data transmission in NB-IoT, we adapt the model and include the link-level results of convolution codes in terms of MCS as shown in Figure \ref{BLER-NBiot}.
In addition, we adapt the number of RBs that is supported by the users for eMTC (1-6 RBs) and NB-IoT (1 RB). We also limit the DL and UL modulation schemes for eMTC to 16QAM and NB-IoT to QPSK. 

Moreover, we dedicate separate subframes for control and data channels. In LTE, both control and data channels are transmitted in the same subframe, while in eMTC and NB-IoT separate subframes are required. We also introduce cross subframe delays and RF switching delays according to 3GPP specifications. In future work, we intend to establish a complete network simulator, implement single-tone NB-IoT and publish the source code with a public license GNU in a public repository.

\section{Simulation results} \label{Sec_Simulation_Result}

In this section, we provide some numerical examples where we evaluate the battery life time of the devices, the latency and the maximum number of devices that can be served in a network.

\subsection{Power consumption}

The power parameters used to evaluate the energy consumption of the devices are based on the data sheet of the ublox SARA-N2 module, where $P_{\text{Tx}}=792$ mW (at transmitted power of 23dBm), $P_{\text{Rx}}= 72$ mW, $P_{\text{Idle}}=22$ mW  and $P_{\text{Sleep}}=18$ $\mu$W. The duration parameters are based on the results of \cite{R1156006}. We assume that the battery capacity is 18000J (5Wh). Hereafter, we do not consider battery leakage impact nor temperature effect nor the peak currents situation. 

\begin{figure}[t!]
  \begin{center}
  \includegraphics[width=1\linewidth]{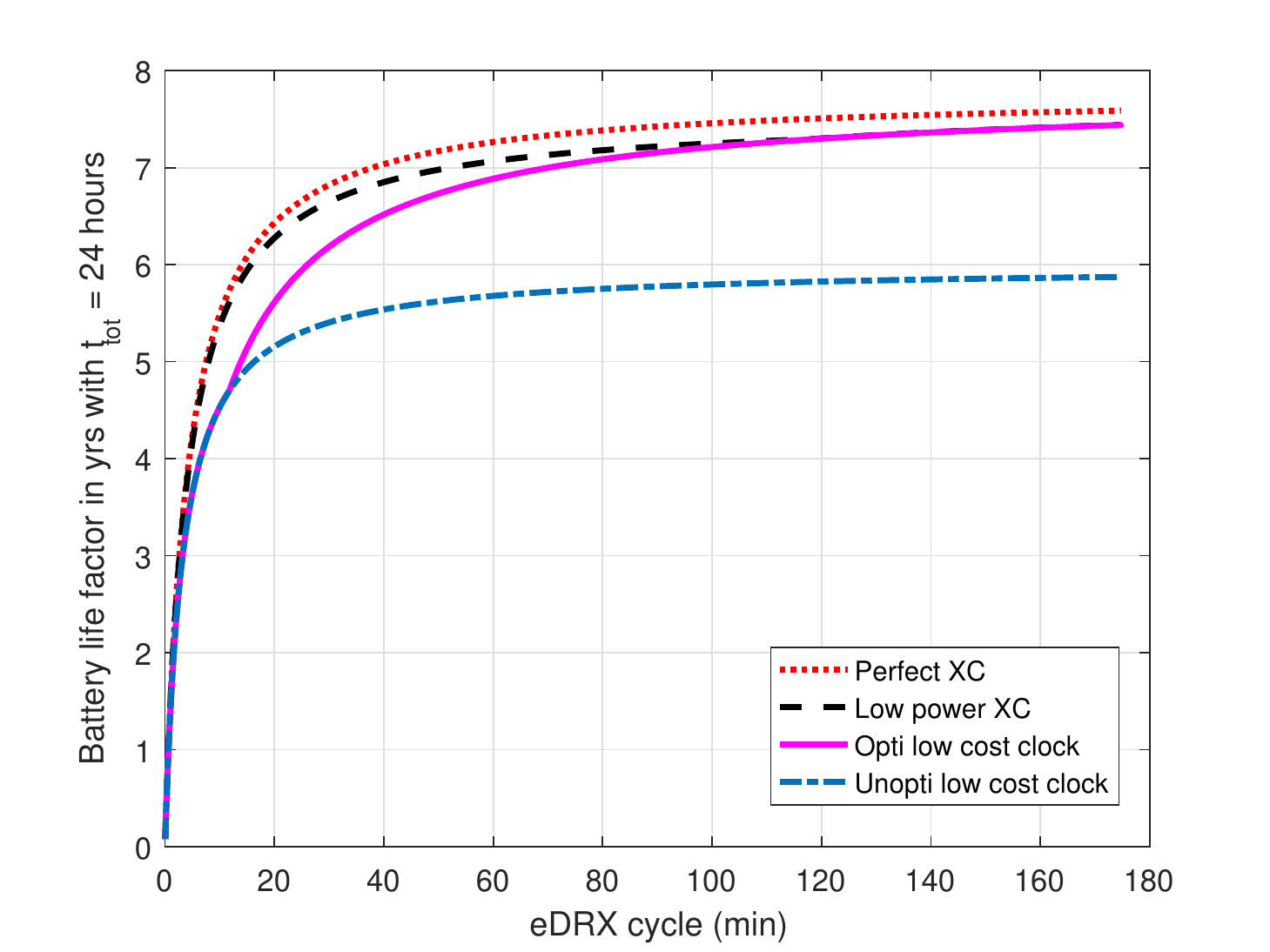}
  \end{center}
  \caption{Battery life time under different eDRX values for NB-IoT at 164dB MCL.}
	\label{DRX164}
\vspace{-.5cm}
\end{figure}

Figure \ref{DRX164} depicts the battery life time obtained for different clocks: perfect crystal clock ``Perfect XC", low power crystal clock ``Low power XC" that has an error of $m=0.01\%$ and optimized using \eqref{OP}, optimized low cost clock ``Opti low cost clock" that has an error of $m=0.1\%$ and optimized using \eqref{OP} and unoptimized low cost clock ``Unopti low cost clock" that has an error of $m=0.1\%$. For the unoptimized low cost clock, the device simply wakes up $m*t_{\text{sleep}}$ earlier. The battery life time is evaluated for NB-IoT at 164dB MCL. 

In Figure \ref{DRX164}, the battery life time is taken as function of eDRX in which we assume that the device is going to transmit a report of 200 bytes with a reporting interval $t_{\text{tot}}$ = 24hrs. We observe that the ``Opti low cost clock" yields more battery life time than the ``Unopti low cost clock" at an eDRX higher than 12min. We also observe that ``Opti low cost clock" yields similar battery life time as ``Low power XC" at low and high eDRX and maximum one year less than the ``Low power XC" at eDRX of 12min. However, the ``Opti low cost clock" is, in general, much cheaper than the ``XC" making the module cheaper and more suitable for IoT. Moreover, it can be easily seen that with a larger sleeping period, i.e. using PSM, the ``Opti low cost clock" yields a battery life time similar to that of ``Low power XC". We should note that in general a low cost clock consumes less energy, however, in this example, we assume that all clocks consume the same energy.

\begin{figure}[t!]
  \begin{center}
  \includegraphics[width=.9\linewidth]{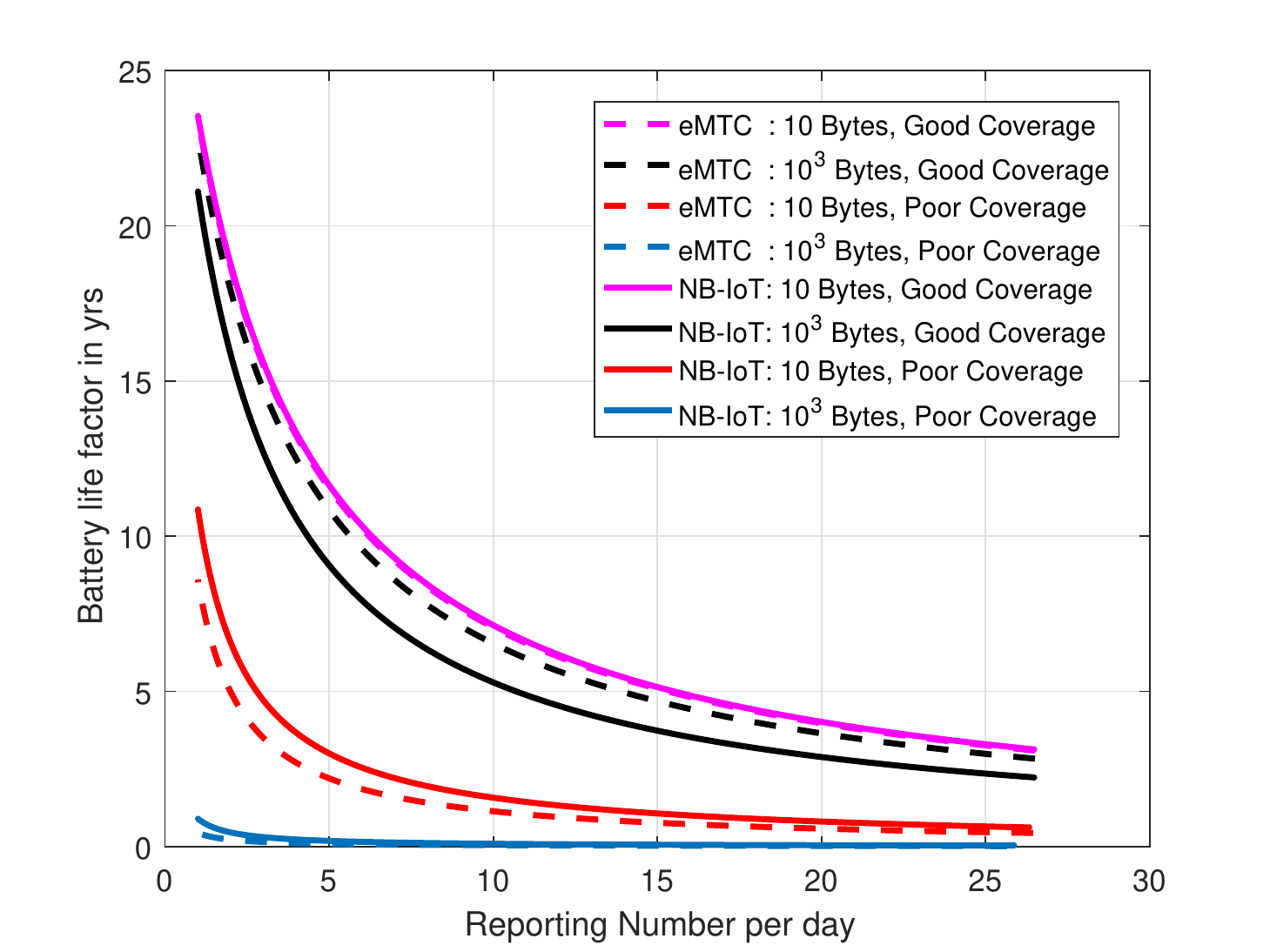}
  \end{center}
  \caption{Battery life time under different reporting number, data lengths and coverage scenarios for eMTC and NB-IoT.}
	\label{eMTCvsNBioT}
\vspace{-.5cm}
\end{figure}

Figure \ref{eMTCvsNBioT} depicts the battery life time obtained using ``Opti low cost clock" under different data lengths and coverage scenarios for eMTC and NB-IoT. The good coverage scenario corresponds to the situation where the node is outdoor (close to the base station) and has the highest MCS value and zero repetition whereas the poor coverage scenario corresponds to the situation where the node is deep indoor (far from the base station) and has the lowest MCS value and the highest repetition value. The battery life time is taken as function of the reporting number per day. We should note that in this figure the PSM technique is being used. We should also note that a bandwidth of 6 RBs and 1 RB is assumed for eMTC and NB-IoT, respectively. We also assume that $P_{\text{Tx}}$ and $P_{\text{Rx}}$ of eMTC are 1.25 times larger than the ones of NB-IoT. 

In Figure \ref{eMTCvsNBioT}, we observe that for small data length NB-IoT has a larger battery life time. We also observe that in a good coverage scenario NB-IoT is slightly higher and in a poor coverage scenario NB-IoT yields a maximum of 2.5 additional years over eMTC. For large data length, we observe that in a good coverage scenario eMTC yields more life time whereas in a poor coverage scenario NB-IoT yields more life time.

In general, the energy consumption or battery life time during a reporting period depends on MCL, data length, bandwidth, RF module and latency. Thus, depending on the values of these parameters, the energy consumption of one technology will be higher or lower than the other. For instance, in a good coverage scenario, even though the RF module of eMTC consumes more energy than the NB-IoT one, eMTC energy consumption is lower if the data length is large. This is due to the larger bandwidth of eMTC compared to NB-IoT, which reduces the transmission time and allows the device to go into sleep mode earlier. However, if the data length is small, the energy consumption of NB-IoT is lower. In a poor coverage scenario, NB-IoT is more efficient, especially in a single-tone transmission. Consequently, its energy consumption is lower than the one of eMTC, which needs a very high number of repetitions. Hence, NB-IoT is a good fit for simple sensors and low-rate applications in medium to poor coverage scenarios, while eMTC is best fit for applications transmitting large amount of data in good to medium coverage scenarios. 

We should note that a battery life time of almost 10 years can be achieved with the current technology in a poor coverage scenario and with a reporting period of 24 hrs. However, for more frequent transmission, e.g., 10 reports per day, the battery life time is decreased to almost one year, which is not practical for IoT modules. Hence, more efforts should be taken to reduce the power consumption of the RF modules, e.g., power amplifier.   

\subsection{Latency and scalability}

\begin{figure}[t!]
  \begin{center}
  \includegraphics[width=.9\linewidth]{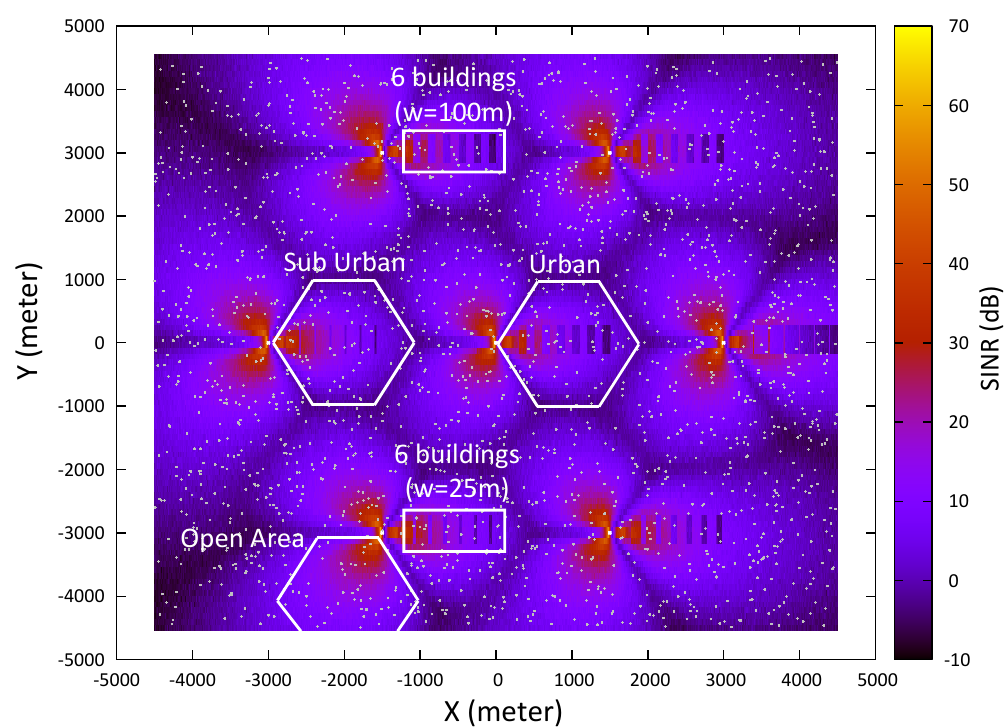}
  \end{center}
  \caption{Radio environment map (REM) of a simple city with urban, suburban and open-area.}
	\label{city-map}
\end{figure}

In order to evaluate the latency and scalability of a network in a smart city, we consider a simple city model that has various types of environments, e.g., urban, suburban and open area, and a varied number of buildings with different inter-building distances as shown in Figure \ref{city-map}. In this city, seven evolved NodeB (eNB) are deployed to serve the UEs, which are placed randomly over the 2D grid and shown by a white point in Figure \ref{city-map}. The figure also shows the signal-to-interference-plus-noise ratio (SINR) values at the UEs.   

Based on this network setup and the NS-3 simulation parameters shown in Table \ref{tab:Simulation-parameters}, we evaluate the latency of the received packets at the cloud and the maximum number of devices that can be served in the different areas. We assume that 92\% of the eNBs bandwidth is allocated to legacy LTE, 6\% to eMTC and 2\% to NB-IoT.

We assume that the number of deployed UEs or devices in each cell can scale up to 1800. In the simulation, we set the propagation model to Hybrid Buildings for indoor and outdoor communications that includes the Hata model, COST231, ITU-R P.1411 (short range communications) and ITU-R P.1238 (indoor communications). All these propagation models are combined in order to be able to evaluate the path loss in different environments (urban, suburban and open area). In addition, we consider an uplink traffic model and we configure each UE to randomly send a packet to the cloud every couple of minutes.

\begin {table}
{\small
\caption {NS-3 simulation parameters} \label{tab:Simulation-parameters} 
\vspace{-.3cm}
\begin{center}
\begin{tabular}{ |p{0.39\columnwidth}|p{0.53\columnwidth}|  }
 \hline
 Parameters & Values\\
 \hline
Number of UEs and eNBs  & 300-1800 per cell and seven eNBs\\
 Propagation model &   Hybrid Buildings \\
 External Wall Loss & Concrete wall no windows (15dB)\\
 Frequency band & DL: 925MHz, UL: 880MHz\\
 System Bandwidth ($\text{N}_{\text{RB}}$) & 6PRBs (eMTC), 2PRBs (NB-IoT) \\
 Uplink traffic period & $<$ 1 minute \\
 Sector per cell & Three  \\
 Antenna model& Parabolic  \\
 Fractional frequency reuse & Hard (FFR) \\ 
 Scheduler & Round Robin\\
 App. packet size &   12 Byte and 160 Byte (IP packet)  \\
 Allocation size & 12 tones @ 15 KHz \\
 Resource Unit (RU) & 1 ms \\
 eNB and UE Tx power & 46 dBm and 20 dBm \\
 OpenArea (good coverage) & 0-33\% indoor,100-67\% outdoor UEs\\
 SubUrban (med. coverage) & 33-66\% indoor,66-33\% outdoor UEs\\
 Urban (poor coverage) & 66-100\% indoor,33-0\% outdoor UEs\\
 \hline
\end{tabular}
\end{center}
}
\vspace{-.7cm}
\end {table}

\begin{figure*}
    \centering
    \begin{subfigure}[b]{0.32\textwidth}
        \includegraphics[width=\textwidth]{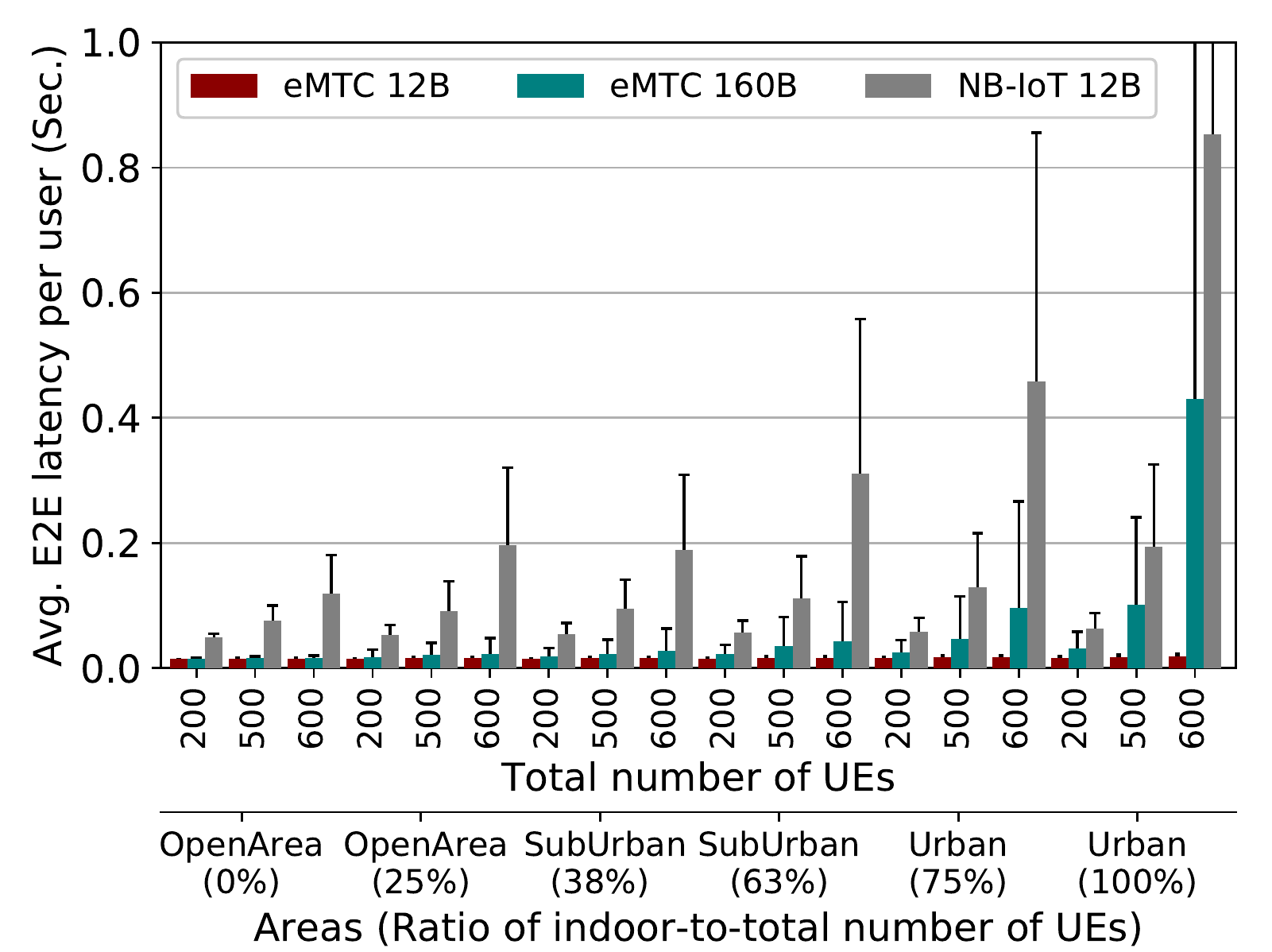}
        \caption{}
        \label{fig_latency}
    \end{subfigure}
    ~ 
    \begin{subfigure}[b]{0.32\textwidth}
        \includegraphics[width=\textwidth]{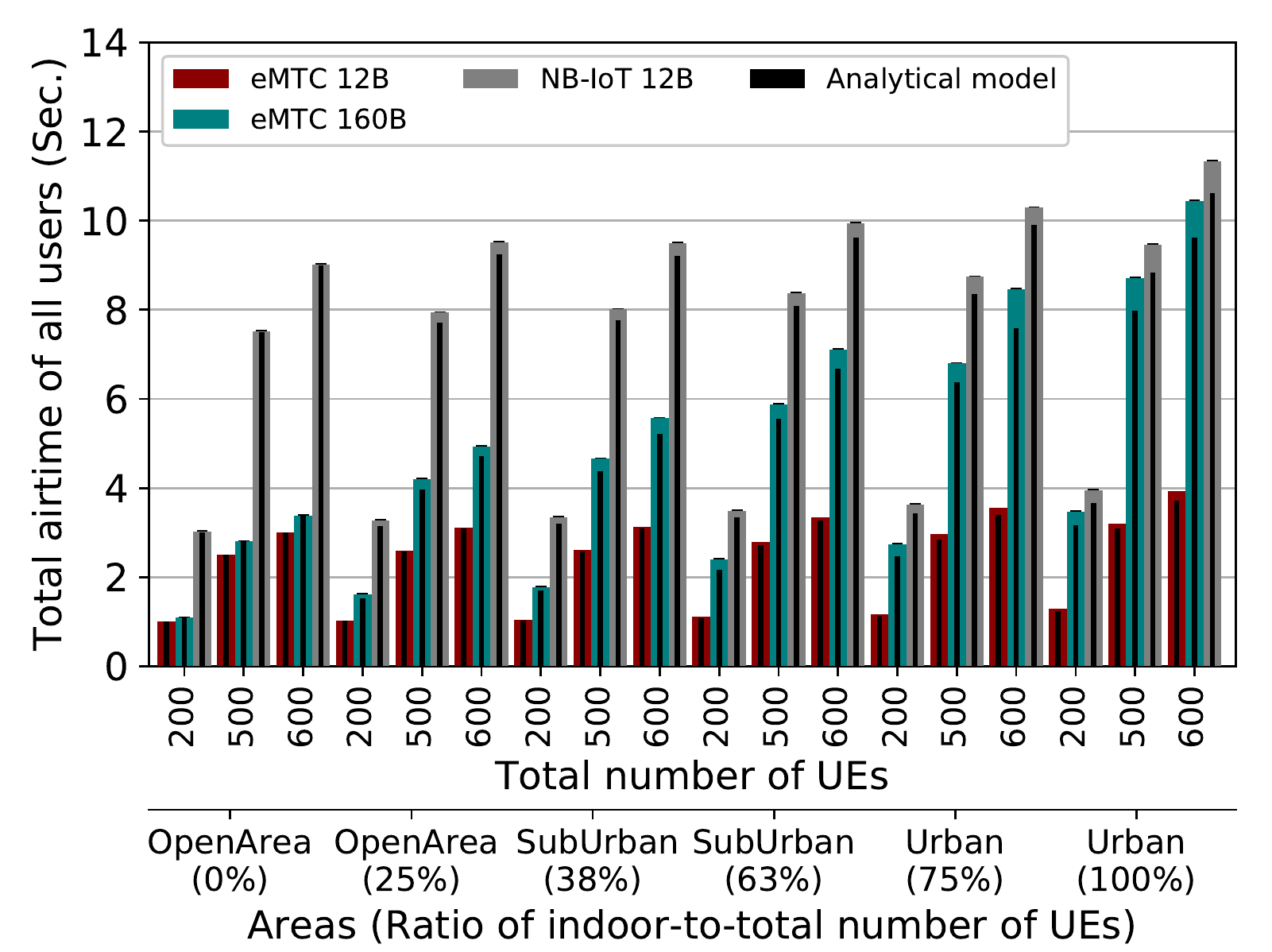}
        \caption{}
        \label{fig_airtime}
    \end{subfigure}
    ~ 
    \begin{subfigure}[b]{0.32\textwidth}
        \includegraphics[width=\textwidth]{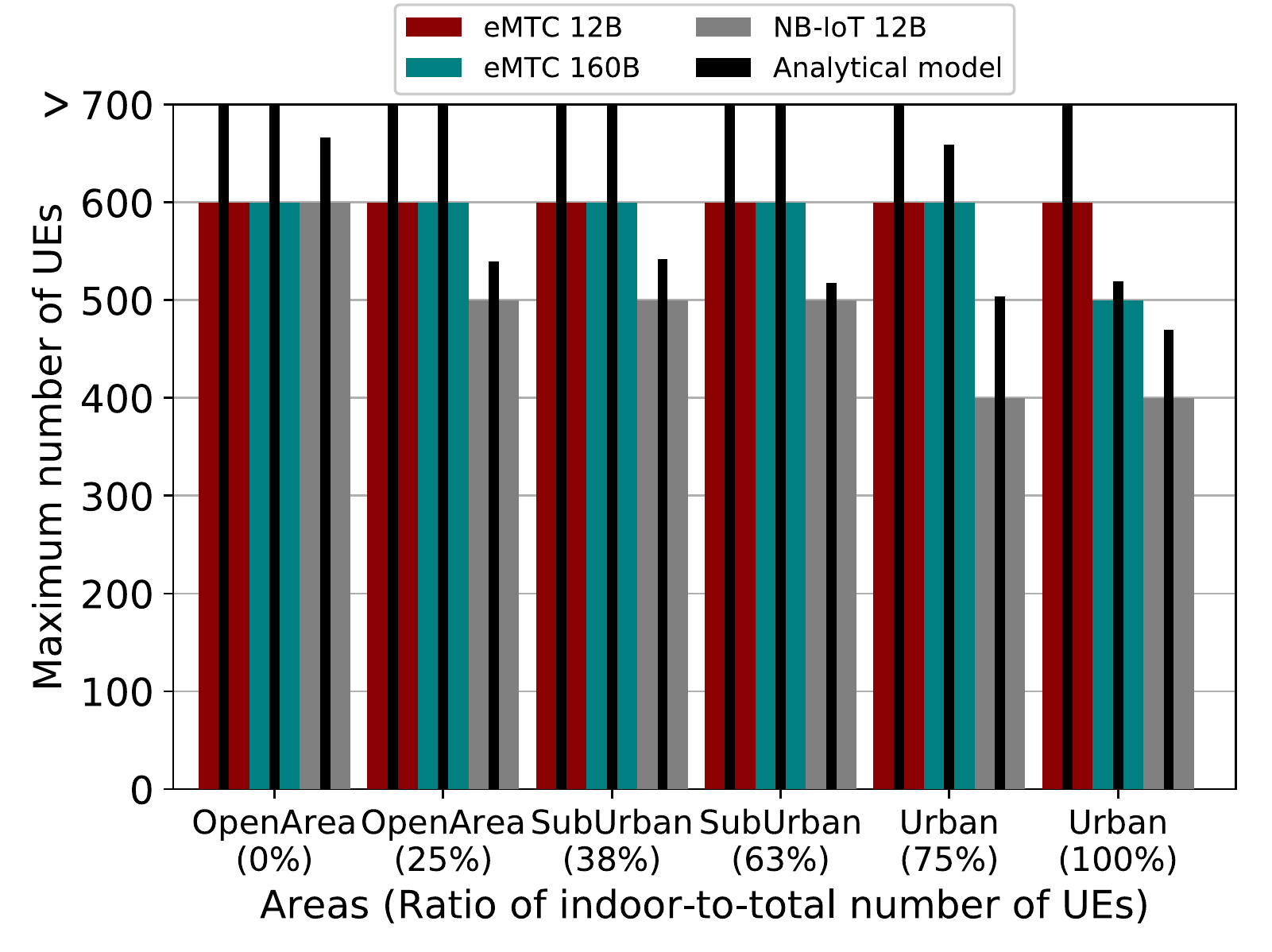}
        \caption{}
        \label{fig_sclalablity}
    \end{subfigure}
    \caption{(a) Average E2E UL latency per user, (b) Total UL air-time of all users, (c) Maximum number of supported users}\label{fig_latency_scalability}
\vspace{-.5cm}
\end{figure*}

Figure \ref{fig_latency} depicts the average end-to-end latency per user on the UL for both eMTC and NB-IoT for a given data length. The latency is taken as function of the ratio of the number of indoor UEs to the total number of available UEs. We assume that all UEs choose a random instant, within the given reporting period, to initiate a transmission. We observe that, for all areas, the delay of transmitting a packet (12 or 160 Bytes) in eMTC is lower than the delay of transmitting a packet (12 Bytes) in NB-IoT. This occurs because eMTC is using a larger bandwidth (6 RBs) than NB-IoT (1 RB), and the cross subframe delays and RF tuning of eMTC are lower than the ones of NB-IoT.   

We also observe that, for all areas, as the number of UEs increases the end-to-end latency per user increases. This occurs because as the number of UEs increases, the probability of waking up and asking for the uplink resources at the same time is higher. The situation aggravates in urban areas (more UEs installed indoor with poor coverage) since the UEs' transmission time increases due to larger MCL. 

Figure \ref{fig_airtime} depicts the total airtime of all users on the UL for both eMTC and NB-IoT for a given data length. It also depicts the total transmission delay obtained in \eqref{eqn:Totaldelay}. We assume that all users wake up at the same instant and simultaneously request uplink resources from eNB. Herein, we use the round robin (RR) scheduler, i.e., the resources are allocated to each user in equal portions and in circular order.  
The total airtime is calculated from the instant the first packet of the first user is transmitted until the last packet of the last user is delivered. The total airtime is taken as function of the ratio of the number of indoor UEs to the total number of available UEs. We observe that the total airtime of all users in eMTC is less than the one in NB-IoT. We also observe that the total data transmission time obtained in \eqref{eqn:Totaldelay}, is similar to the results obtained in NS-3. We should note that in indoor scenarios more repetitions and smaller TBS are required, which explains the higher total air-time. 

Figure \ref{fig_sclalablity} depicts the maximum number of UEs that can be supported in different environments of a network. It also depicts the maximum number of UEs obtained in \eqref{eqn:maxNumberUEs}. We assume that all UEs choose a random instant, within the given reporting period, to initiate a transmission. We should note that this figure only displays the number of UEs who have more than 90\% successfully delivered packets at the eNB/cloud in the given reporting period. Due to memory and process limitation in the simulation, the maximum number of UEs that can be simulated is limited to 600. We should also note that the maximum number of UEs is constrained by the reporting period and the percentage of indoor UEs. We can observe in Figure \ref{fig_sclalablity} that eMTC can support more users than NB-IoT. For instance, eMTC at 100\% of indoor UEs, can serve 500 UEs (with a 160 Bytes data length) while for NB-IoT this is not the case. In the latter, when the number of indoor UEs is above 400, the airtime is high and the eNB/scheduler, that has only 2 RBs, cannot guarantee anymore the delivery of UEs' reports in a timely manner. Thus, the request of some UEs is discarded. However, in eMTC, since the airtime is low, the eNB/scheduler, that has 6 RBs, can deliver all 500 UEs' reports in a timely manner. Moreover, we can observe that the maximum number of users obtained in \eqref{eqn:maxNumberUEs} also decreases whenever there are more indoor users than outdoor. We should note that the results obtained using the analytical model are slightly larger than the simulated ones. This is because in the analytical model, unlike in NS-3, we assume that there are no collisions among the users requests and that the reporting period is optimally divided among the UEs.

\section{Conclusion}
In this paper, we study two recent LPWAN technologies, eMTC and NB-IoT, that are based on LTE and we evaluate their performance for smart city applications. We evaluate the battery life time of both technologies considering the clock drift. We show that a battery life time of 8 years can be achieved by both technologies in a poor coverage scenario with a reporting interval of one day. We also formulate the data transmission delay and the maximum number of UEs per cell or network and show that these metrics depend on MCL, channel bandwidth, data length, RF switching delay, cross subframe delay as well as the reporting period. In addition, we introduce a network simulator tool that can evaluate the end-to-end latency and scalability of eMTC and NB-IoT in a large-scale network in various environments. We show that eMTC can serve more devices in a network than NB-IoT, while providing a latency that is 10 times lower. We also demonstrate that the results obtained from the network simulator correspond to the results derived from the analytical formulations. In future works, we intend to compare the simulator results of eMTC and NB-IoT in NS-3 with field-test results.



%



\ifCLASSOPTIONcaptionsoff
  \newpage
\fi



%

\bibliographystyle{IEEEtran}

\begin{thebibliography}{10}
\providecommand{\url}[1]{#1}
\csname url@samestyle\endcsname
\providecommand{\newblock}{\relax}
\providecommand{\bibinfo}[2]{#2}
\providecommand{\BIBentrySTDinterwordspacing}{\spaceskip=0pt\relax}
\providecommand{\BIBentryALTinterwordstretchfactor}{4}
\providecommand{\BIBentryALTinterwordspacing}{\spaceskip=\fontdimen2\font plus
\BIBentryALTinterwordstretchfactor\fontdimen3\font minus
  \fontdimen4\font\relax}
\providecommand{\BIBforeignlanguage}[2]{{%
\expandafter\ifx\csname l@#1\endcsname\relax
\typeout{** WARNING: IEEEtran.bst: No hyphenation pattern has been}%
\typeout{** loaded for the language `#1'. Using the pattern for}%
\typeout{** the default language instead.}%
\else
\language=\csname l@#1\endcsname
\fi
#2}}
\providecommand{\BIBdecl}{\relax}
\BIBdecl

\bibitem{Intel17}
``A guide to the internet of things,'' Available at
  \url{https://www.intel.com/content/www/us/en/internet-of-things/infographics/guide-to-iot.html}.

\bibitem{LEIOT17}
``{LTE} evolution for {IoT} connectivity,'' Available at
  \url{http://resources.alcatel-lucent.com/asset/200178} (2017/01/01).

\bibitem{RKS17}
U.~Raza, P.~Kulkarni, and M.~Sooriyabandara, ``Low power wide area networks: An
  overview,'' \emph{IEEE Communications Surveys Tutorials}, vol.~19, no.~2, pp.
  855--873, Second quarter 2017.

\bibitem{TS36300}
``{Evolved Universal Terrestrial Radio Access (E-UTRA) and Evolved Universal
  Terrestrial Radio Access Network (E-UTRAN); Overall description},'' {3GPP TS
  36.300}, 2016.

\bibitem{TR45820}
``{Cellular System Support for ultra-low complexity and low throughput Internet
  of Things (CIoT)},'' {TR 45.820}, 2015.

\bibitem{TS23682}
``{Architecture enhancements to facilitate communications with packet data
  networks and applications},'' {3GPP TS 23.682}, 2016.

\bibitem{LKMSH16}
M.~Lauridsen, I.~Z. Kovacs, P.~Mogensen, M.~Sorensen, and S.~Holst, ``Coverage
  and capacity analysis of lte-m and nb-iot in a rural area,'' in \emph{2016
  IEEE 84th Vehicular Technology Conference (VTC-Fall)}, Sept 2016, pp. 1--5.

\bibitem{RVMG16}
R.~Ratasuk, B.~Vejlgaard, N.~Mangalvedhe, and A.~Ghosh, ``Nb-iot system for m2m
  communication,'' in \emph{2016 IEEE Wireless Communications and Networking
  Conference}, April 2016, pp. 1--5.

\bibitem{R1156006}
R1-156006, ``{NB-IoT - battery lifetime evaluation},'' {3GPP TSG RAN1\#82bis,
  2015}.

\bibitem{RVXWBBTY16}
A.~Rico-Alvarino, M.~Vajapeyam, H.~Xu, X.~Wang, Y.~Blankenship, J.~Bergman,
  T.~Tirronen, and E.~Yavuz, ``An overview of 3gpp enhancements on machine to
  machine communications,'' \emph{IEEE Communications Magazine}, vol.~54,
  no.~6, pp. 14--21, June 2016.

\bibitem{RMZRK16}
R.~Ratasuk, N.~Mangalvedhe, Y.~Zhang, M.~Robert, and J.~P. Koskinen, ``Overview
  of narrowband iot in lte rel-13,'' in \emph{2016 IEEE Conference on Standards
  for Communications and Networking (CSCN)}, Oct 2016, pp. 1--7.

\bibitem{WLAGSBBR17}
Y.~P.~E. Wang, X.~Lin, A.~Adhikary, A.~Grovlen, Y.~Sui, Y.~Blankenship,
  J.~Bergman, and H.~S. Razaghi, ``A primer on 3gpp narrowband internet of
  things,'' \emph{IEEE Communications Magazine}, vol.~55, no.~3, pp. 117--123,
  March 2017.

\bibitem{MMRBZ12}
M.~Mezzavilla, M.~Miozzo, M.~Rossi, N.~Baldo, and M.~Zorzi, ``A lightweight and
  accurate link abstraction model for the simulation of {LTE} networks in
  ns-3,'' in \emph{Proceedings of the 15th ACM International Conference on
  Modeling, Analysis and Simulation of Wireless and Mobile Systems}.\hskip 1em
  plus 0.5em minus 0.4em\relax New York, NY, USA: ACM, 2012, pp. 55--60.

\end{thebibliography}

%

%
%
%




\end{document}